\documentclass[aps,pre,preprint,groupedaddress]{revtex4}
\usepackage{graphics,epsfig}
\usepackage{amsmath,amssymb}

\begin{document}

\title{Transitions in collective response in multi-agent models of
competing populations driven by resource level}

\author{Sonic H. Y. Chan,$^1$ T. S. Lo,$^{1}$ P. M. Hui,$^1$ and
N. F. Johnson$^2$}

\affiliation{$^{1}$Department of Physics, The
Chinese University of Hong Kong\\
Shatin, New Territories, Hong Kong\\
$^{2}$Department of Physics, University of Oxford, Oxford OX1 3PU,
United Kingdom}

\begin{abstract}
We aim to study the effects of controlling the resource level in
agent-based models.  We study, both numerical and analytically, a
Binary-Agent-Resource (B-A-R) model in which $N$ agents are
competing for resources described by a resource level $1/2 \leq
{\cal L} < 1$, where ${\cal L} = L/N$ with $L$ being the maximum
amount of resource per turn available to the agents.  Each agent
picks the momentarily best-performing strategy for decision with
the performance of the strategy being a result of the cumulative
collective decisions of the agents. The agents may or may not be
networked for information sharing.  Detailed numerical simulations
reveal that the system exhibits well-defined plateaux regions in
the success rate which are separated from each other by abrupt
transitions.  As $L$ increases, the maximum success rate forms a
well defined sequence of simple fractions.  We analyze the
features by studying the outcome time series, the dynamics of the
strategies' performance ranking pattern and the dynamics in the
history space.  While the system tends to explore the whole
history space due to its competitive nature, an increasing $L$ has
the effect of driving the system to a restricted portion of the
history space.  Thus the underlying cause of the observed features
is an interesting self-organized phenomena in which the system, in
response to the global resource level, effectively avoids
particular patterns of history outcomes.  We also compare results
in networked population with those in non-networked population.

\noindent {\bf Paper to be presented in the 10th Annual Workshop
on Economic Heterogeneous Interacting Agents (WEHIA 2005), 13-15
June 2005, University of Essex, UK.}

\end{abstract}

\maketitle \thispagestyle{empty}

\section{Introduction}
\label{sec:Intro}

Looking around the world that we live in, we will find that there
are complex systems everywhere. The study of these systems has
become a hot research area in different disciplines such as
physics, applied mathematics, biology, engineering, economics, and
social sciences.  In particular, agent-based models have become an
essential part of research on Complex Adaptive Systems (CAS)
\cite{recentactivities}.  For example, self-organized phenomena in
an evolving population consisting of agents competing for a
limited resource, have potential applications in areas such as
engineering, economics, biology, and social sciences
\cite{recentactivities,Johnson2003a}. The famous El Farol bar
attendance problem proposed by Arthur
\cite{Arthur1994a,Johnson1998a} constitutes a typical example of
such a system in which a population of agents decide whether to go
to a popular bar having limited seating capacity.  The agents are
informed of the attendance in the past weeks, and hence share
common information, make decisions based on past experience,
interact through their actions, and in turn generate this common
information collectively.  These ingredients are the key
characteristics of complex systems \cite{Johnson2003a}.  The
proposals of the binary versions of models of competing
populations, either in the form of the Minority Game (MG)
\cite{Challet1997a} or in a Binary-Agent-Resource (B-A-R) game
\cite{Johnson2003b,Johnson1999a}, have led to a deeper
understanding in the research in agent-based models.  For modest
resource levels in which there are more losers than winners, the
Minority Game proposed by Challet and Zhang
\cite{Challet1997a,Challet2003a} represents a simple, yet
non-trivial, model that captures many of the essential features of
such a competing population.  The MG, suitably modified, can be
used to model financial markets and reproduce the stylized facts.
The B-A-R model, which is a more general model in which the
resource level is taken to be a parameter, has much richer
behaviour.  In particular, we will discuss the model and report
the emergence of plateaux-and-jump structures in the average
success rates in the population as the resource level is varied.
We analyze the results within the ideas of the trail of histories
in the history space and the strategy performance ranking
patterns.

\section{Model}
\label{sec:Model}

The binary-agent-resource (B-A-R) model
\cite{Johnson2003b,Johnson2004a,Johnson1999a} is a binary version
of Arthur's El Farol bar attendance model
\cite{Arthur1994a,Johnson1998a}, in which a population of $N$
agents repeatedly decide whether to go to a bar with limited
seating based on the information of the crowd size in recent
weeks.  In the B-A-R model, there is a global resource level $L$
which is not announced to the agents.  At each timestep $t$, each
agent decides upon two possible options: whether to access
resource $L$ (action `1') or not (action `0').  The two global
outcomes at each timestep, `resource overused' and `resource not
overused', are denoted by $0$ and $1$.  If the number of agents
$n_{1}(t)$ choosing action `$1$' exceeds $L$ (i.e., resource
overused and hence global outcome $0$) then the $N - n_{1}(t)$
abstaining agents win.  By contrast if $n_{1}(t) \leq L$ (i.e.,
resource not overused and hence global outcome $1$) then the
$n_{1}(t)$ agents win.  In order to investigate the behaviour of
the system as $L$ changes, it is sufficient to study the range
$N/2 \leq L \leq N$.  The results for the range $0 \leq L \leq
N/2$ can be obtained from those in the present work by suitably
interchanging the role of `0' and `1' \cite{Johnson1999a}. In the
special case of $L = N/2$, the B-A-R model reduces to the Minority
Game.

In the B-A-R model, each agent shares a common knowledge of the
past history of the most recent $m$ outcomes, i.e., the winning
option in the most recent $m$ timesteps.  The agents are essential
identical, except for the details in the strategies that they are
holding.  The full strategy space thus consists of $2^{2^m}$
strategies, as in the MG. Initially, each agent randomly picks $s$
strategies from the pool of strategies, with repetitions allowed.
The agents use these strategies throughout the game.  At each
timestep, each agent uses his momentarily best performing strategy
with the highest virtual points.  The virtual points for each
strategy indicate the cumulative performance of that strategy: at
each timestep, one \textbf{virtual point} (VP) is awarded to a
strategy that would have predicted the correct outcome after all
decisions have been made, whereas it is unaltered for those with
incorrect predictions.  Notice that in the literature, sometimes
one VP is deducted for an incorrect prediction.  The results
reported here, however, come out to be the same.  A random
coin-toss is used to break ties between strategies. In the B-A-R
model, the agents in the population may or may not be connected,
by some kind of network. In the case of a networked population
\cite{Gourley2004a,my2004b,Anghel2004a,Choe2004a,keven2005}, each
agent has access to additional information from his connected
neighbours, such as his neighbours' strategies and/or
performance. Here, we will focus our discussion on the B-A-R
model in a non-networked population and report some numerical
results for networked population.

The B-A-R model thus represents a general setting of a competing
population in which the resource level can be tuned.  From a
governmental management point of view, for example, one would like
to study how a population may react to a decision on increasing or
decreasing a certain resource in a community.  Will such a change
in resource level lead to a large response in the community or the
community will be rather insensitive?  To evaluate the performance
of an agent, one \textbf{(real) point} is awarded to each winning
agent at a given timestep.  A maximum of $L$ points per turn can
therefore be awarded to the agents per timestep.  An agent has a
success rate $w$, which is the mean number of points awarded to
the agent per turn over a long time window. The mean success rate
$\langle w \rangle$ among the agents is then defined to be the
mean number of points awarded per agent per turn, i.e., an average
of $w$ over the agents.  We are interested in investigating the
details of how the success rates, including the mean success rate
and the highest success rate among the agents, change as the
resource level $L$ varies in the efficient phase, where the number
of strategies (repetitions counted) in play is larger than the
total number of distinct strategies in the strategy space.

\section{Results: Plateaux formation and abrupt jumps}
\label{sec:Phases}

The effects of varying $L$ were first studied by Johnson {\em et
al.} \cite{Johnson1999a}.  These authors reported the dependence
of the fluctuations in the number of agents taking a particular
option, on the memory size $m$ for different values of $L$.  For
the MG (i.e., $L = N/2$) in the efficient phase (i.e., small
values of $m$) the number of agents making a particular choice
varies from timestep to timestep, with additional stochasticity
introduced via the random tie-breaking process.  The corresponding
period depends on the memory length $m$.  The underlying reason is
that in the efficient phase for $L=N/2$, no strategy is better
overall than any other.  Hence there is a tendency for the system
to restore itself after a finite number of timesteps, thereby
preventing a given strategy's VPs from running away from the
others.  As a result, the outcome bit-string shows the feature of
anti-persistency or double periodicity
\cite{Challet2000a,Challet2001a,Challet1999a,Marsili2000a,Savit1999a,Jefferies2002a,Zheng2001a}.
Since a maximum of $L=N/2$ points can be awarded per turn, the
mean success rate $\langle w \rangle$ over a sufficiently large
number of timesteps is bound from above by $L/N = 1/2$.

\begin{figure}[!htbp]
\centerline{\epsfig{file=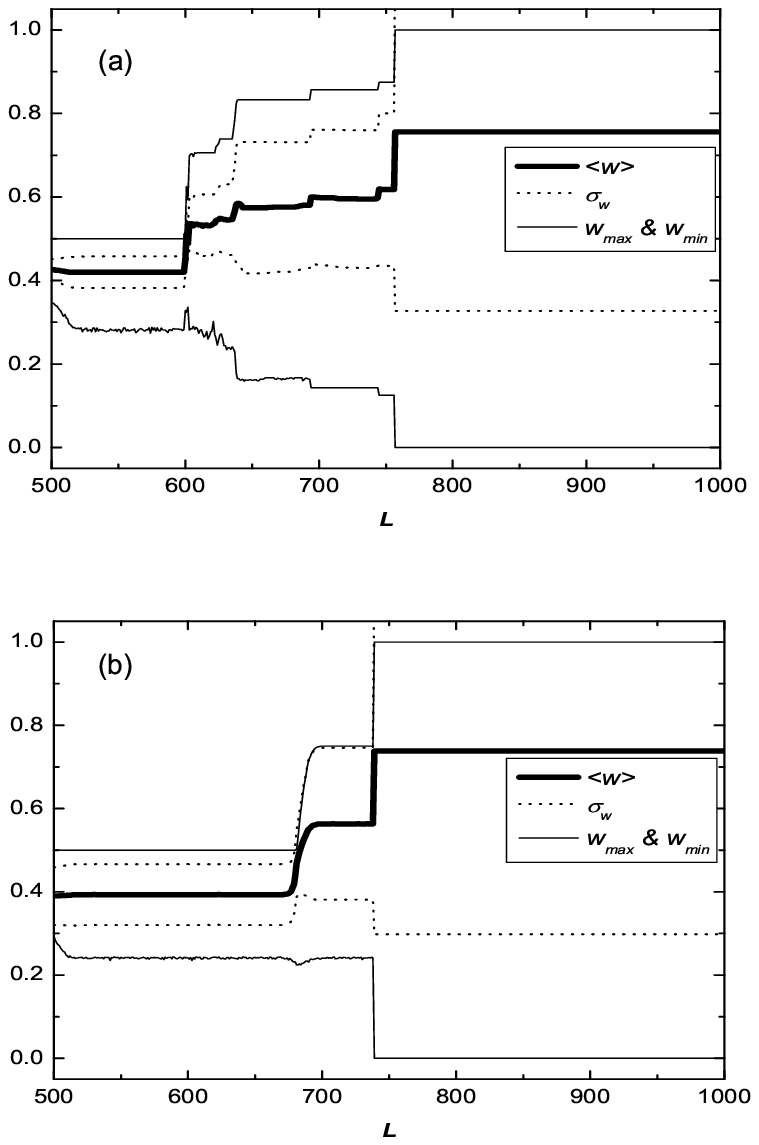,height=7in}} \caption[Statistics of
the success rate $w$ as a function of resource level $L$ with $N=1001$,
$s=2$, (a) $m=3$ and (b) $m=1$.]{Statistics of the success rate $w$ as a
function of resource level $L$ with $N=1001$, $s=2$, (a) $m=3$ and (b)
$m=1$. For each value of $m$, the agents have the same set of strategies for
all $L$. } \label{fig:w_m1m3s2}
\end{figure}

We have carried out extensive numerical simulations on the B-A-R
model to investigate the dependence of the success rate on $L$ for
$N/2 \leq L \leq N$.  Unless stated otherwise, we consider systems
with $N=1001$ agents and $s=2$.  Figure~\ref{fig:w_m1m3s2} shows
the results of the mean success rate ($\langle w \rangle$, dark
solid line) as a function of $L$ in a typical run for (a) $m=3$
and (b) $m=1$, together with the range corresponding to one
standard deviation about $\langle w \rangle$ in the success rates
among the $N$ agents ($\sigma_w$, dotted lines) and the spread in
the success rates given by the highest and the lowest success
rates ($w_{\max}$ and $w_{\min}$, thin solid lines) in the
population.

By taking a larger value of $N$ than most studies in the
literature, we are able to analyze the dependence on $L$ and $m$
in greater detail and discover new features.  In particular, these
quantities all exhibit {\em abrupt transitions} (i.e., jumps) at
particular values of $L$. Between the jumps, the quantities remain
essentially constant and hence form steps or `plateaux'.  We refer
to these different plateaux as states or phases, since it turns
out that the jump occurs when the system makes a transition from
one type of state characterizing the outcome bit-string to
another.  The origins of the plateaux are due to (i) the finite
number of states allowed in each $m$, and (ii) insensitive to $L$
inside each state.  The success rates within each plateau (state)
come out to the the same.  For different runs, the results are
almost identical.  At most, there are tiny shifts in the $L$
values at which jumps arise due to (i) different initial strategy
distributions among the agents in different runs, and (ii)
different random initial history bit-strings used to start the
runs.  In other words, a uniform distribution of strategies among
the agents (e.g., in the limit of a large population) gives stable
values of $L$ at the jumps. The results indicate that a community
does not react to a slight change in resource level that lies
within a certain plateau, but the response will be abrupt if the
change in resource level swaps through critical values between
transitions.

\begin{table}[!htb]
\begin{center}
\begin{tabular}{|c||c|c|c|}
\hline
 $w_{\max}$ & 1-bits:0-bits & Range of $L$ & \phantom{M}Period\phantom{M} \\
\hline \hline
 $1/2$ & 8:8 & $\sim$ 510-600 & length 16 \\
\hline
 $12/17$ & 12:5 & $\sim$ 600-620 & length 17 \\
\hline
 $17/23$ & 17:6 & $\sim$ 620-640 & length 23 \\
\hline
 $5/6$ & 5:1 & $\sim$ 640-695 & 111110 \\
\hline
 $6/7$ & 6:1 & $\sim$ 695-745 & 1111110 \\
\hline
 $7/8$ & 7:1 & $\sim$ 745-755 & 11111110 \\
\hline
 $1$ & 1:0 & $\sim$ 755-1000 & 1 \\
\hline
\end{tabular}
\end{center}
\caption[The states for the B-A-R model with $N=1001$, $m=3$ and
$s=2$.]{The states characterized by $w_{\max}$ for the B-A-R model
with $N=1001$, $m=3$ and $s=2$.  The results are obtained from the
numerical data shown in Fig.~\ref{fig:w_m1m3s2}(a). }
\label{tab:ots_m3s2}
\end{table}

\begin{table}[!htb]
\begin{center}
\begin{tabular}{|c||c|c|c|}
\hline
 $w_{\max}$ & 1-bits:0-bits & Range of $L$ & \phantom{M}Period\phantom{M} \\
\hline \hline
 $1/2$ & 2:2 & $\sim$ 520-650 & 1100 \\
\hline
 $3/4$ & 3:1 & $\sim$ 680-750 & 1110 \\
\hline
 $1$ & 1:0 & $\sim$ 750-1000 & 1 \\
\hline
\end{tabular}
\end{center}
\caption[The states for the B-A-R model with $N=1001$, $m=1$ and
$s=2$.]{The states characterized by $w_{\max}$ for the B-A-R model
with $N=1001$, $m=1$ and $s=2$. The results are obtained from the
numerical data shown in Fig.~\ref{fig:w_m1m3s2}(b). }
\label{tab:ots_m1s2}
\end{table}

The most striking feature in Fig.~\ref{fig:w_m1m3s2}(a) is that
the values of the plateaux in the highest success rate $w_{\max}$,
are given by {\em simple fractions}, e.g., $1$, $7/8$, $6/7$,
$5/6$, $12/17$, $1/2$, etc.  Figure~\ref{fig:w_m1m3s2}(b) shows
that the features in the success rates for the simpler case of
$m=1$ are similar to those in Fig.~\ref{fig:w_m1m3s2}(a), except
that the plateaux in $w_{\max}$ take on fewer values, i.e., $1$,
$3/4$, $1/2$ as $L$ decreases.  These values are closely related
to the statistics in the outcome bit-string.  For large $N$ and
$m=1$, the outcome bit-string shows a period of 4 bits. For values
of $L$ with $w_{\max} = 1/2$, it turns out that the fraction of
the outcome $1$ in a period is exactly $1/2$.  For the range of
$L$ corresponding to $w_{\max} = 3/4$, there are three $1$'s in a
period of $4$.  At the range corresponding to $w_{\max}=1$, there
are persistently 1. Thus, the system passes through different
states with different ratios of 1 and 0 in the outcome bit-string
as $L$ varies.  For $m=3$, we have also carried out detailed
analysis of the outcome bit-string. For later discussions, we
summarize in Tables~\ref{tab:ots_m3s2} and \ref{tab:ots_m1s2} the
values of $w_{\max}$, the range of $L$ corresponding to the
observed value of $w_{\max}$, the ratio of number of occurrence of
1-bits to 0-bits and the period in the outcome bit-string, as
obtained numerically from the data shown in
Fig.~\ref{fig:w_m1m3s2}. Hereafter, $w_{\max}$ is used to label
the state at a given $L$.

\section{Why is it so? The history space and Bit-string patterns}
\label{sec:Macro}

\subsection{The history space}
\label{ssec:Macro_History}

\begin{figure}[!htb]
\centerline{\psfig{file=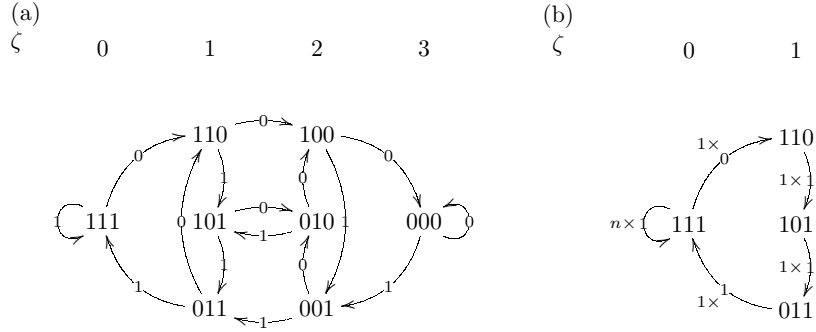,height=2in}} \caption[The
history space in the B-A-R model with $m=3$.  (a) The complete
history space.  (b) The path for $\zeta_{\max}=1$ states.]{(a) The
history space in the B-A-R model with $m=3$. The nodes correspond
to the $2^{m}$ possible histories.  The transition between nodes
are indicated by the arrows, together with the outcome needed for
the transitions to occur.  The histories can be grouped into
columns labeled by a parameter $\zeta$ which gives the number of
`0' bits in the histories. (b) For the $\zeta_{\max}=1$ states at
$L \lesssim (1- 1/2^{s})N$, the system follows a path restricted
to histories in the $\zeta=0$ and $\zeta=1$ columns, with $n$
loops at the $\zeta=0$ history.} \label{fig:hs_m3}
\end{figure}

We analyze the underlying mechanism for the emergence of plateaux
and jumps in the performance of the system in terms of the path
that a system passes through in the history space and the strategy
performance ranking as the system evolves \cite{my2004b}. The
history space consists of all the possible history bit-strings for
a given value of $m$. For $m=3$, it includes $2^{3}$ bit-strings
of $0$'s and $1$'s. Figure~\ref{fig:hs_m3}(a) shows the history
space for $m=3$, together with the possible transitions (arrows)
from one history to another.  It will prove convenient to group
the possible history bit-strings for a given $m$ into columns, in
the way shown in Fig.~\ref{fig:hs_m3}(a). We label each column by
a parameter $\zeta$, which is the number of $0$'s in the $3$-bit
history (histories) concerned. One immediate advantage of this
labelling scheme is that the different states characterized by
$w_{\max}$ turn out to involve paths in a {\em restricted} portion
of the full history space.  For example, the state with
$w_{\max}=1$ is restricted to the $\zeta = 0$ portion of the
history space, i.e., the $111$ history bit-string leads to an
outcome of 1 and hence persistent self-looping at the node $111$
in history space.  The states with $w_{\max}=7/8$, $6/7$, and
$5/6$ correspond to different paths in the history space
restricted to the $\zeta=0$ and $\zeta=1$ groups of histories, as
shown in Fig.~\ref{fig:hs_m3}(b).  The states with
$w_{\max}=17/23$ and $12/17$ have paths extended to include
$\zeta=2$ histories.  The state with $w_{\max}=1/2$ has paths that
visit the whole history space ($\zeta = 0,1,2,3$).  The
classification is summarized in Table~\ref{tab:class_m3}.  One can
regard this behaviour as the system, in response to the global
resource level $L$, effectively avoiding certain nodes in the
history space and hence avoiding particular patterns of historical
outcomes.  Thus, the resource level may be used as a control on
the population's response, in a way that restricts the agents from
using every part of their strategies by suppressing the
occurrence of some of the history bit-strings.

\begin{table}[!htb]
\begin{center}
\begin{tabular}{ccl}
\hline
 \phantom{M}$\zeta_{\max}$\phantom{M} & \phantom{M} & States\phantom{M} \\
\hline \hline
 3 & & $\frac{8}{16}$ \\
\hline
 2 & & $\frac{12}{17}, \frac{17}{23}$ \\
\hline
 1 & & $\frac{5}{6}, \frac{6}{7}, \frac{7}{8}$ \\
\hline
 0 & & $1$ \\
\hline
\end{tabular}
\end{center}
\caption[Classification of states by means of column number
$\zeta$ for $m=3$.]{Classification of states by means of column
number $\zeta$ for $m=3$.  The states characterized by $w_{\max}$
have paths in the history space covering nodes in the columns
$\zeta = 0, \cdots, \zeta_{\max}$.} \label{tab:class_m3}
\end{table}

\subsection{Outcome Bit-string statistics at different resource levels}
\label{ssec:z0states}

As the game proceeds, the system evolves from one history
bit-string to another.  This can be regarded as transitions
between different nodes (i.e., different histories) in the history
space.  For $L=N/2$ in the efficient phase, it has been shown
\cite{Savit1999a} that the conditional probability of an outcome
of, say, $1$ following a given history is the same for all
histories.  For $L \neq N/2$, the result still holds for states
characterized by $w_{\max}=1/2$.  Note that a history bit-string
can only make transitions to history bit-strings that differ by
the most recent outcome, e.g., 111 can only be make transitions to
either 110 or 111, and thus many transitions between two chosen
nodes in the history space are forbidden.  In addition, these
allowed transitions do not in general occur with equal
probabilities. This leads to specific outcome (and history)
bit-string statistics for a state characterized by $w_{\max}$.

\begin{table}[!htb]
\setlength{\tabcolsep}{0.5mm}
\begin{center}
\begin{tabular}{|c|c|cc|}
\hline
 \multicolumn{2}{|c|}{$w_{\max}=\frac{8}{16}$} & $\rightarrow$0 & $\rightarrow$1 \\
\hline \hline
 $\zeta=3$ & 000 & 1 & 1 \\
\hline
           & 001 & 1 & 1 \\
 $\zeta=2$ & 010 & 1 & 1 \\
           & 100 & 1 & 1 \\
\hline
           & 011 & 1 & 1 \\
 $\zeta=1$ & 101 & 1 & 1 \\
           & 110 & 1 & 1 \\
\hline
 $\zeta=0$ & 111 & 1 & 1 \\
\cline{1-4} \cline{3-4}
 \multicolumn{2}{c|}{} & 8 & 8 \\
\cline{3-4}
\end{tabular}
\begin{tabular}{|c|cc|}
\hline
 $\frac{12}{17}$ & $\rightarrow$0 & $\rightarrow$1 \\
\hline \hline
 000 & 0 & 0 \\
\hline
 001 & 0 & 1 \\
 010 & 0 & 1 \\
 100 & 0 & 1 \\
\hline
 011 & 1 & 2 \\
 101 & 1 & 2 \\
 110 & 1 & 2 \\
\hline
 111 & 2 & 3 \\
\cline{1-3} \cline{2-3}
 \multicolumn{1}{c|}{} & 5 & 12 \\
\cline{2-3}
\end{tabular}
\begin{tabular}{|c|cc|}
\hline
 $\frac{17}{23}$ & $\rightarrow$0 & $\rightarrow$1 \\
\hline \hline
 000 & 0 & 0 \\
\hline
 001 & 0 & 1 \\
 010 & 0 & 1 \\
 100 & 0 & 1 \\
\hline
 011 & 1 & 3 \\
 101 & 1 & 3 \\
 110 & 1 & 3 \\
\hline
 111 & 3 & 5 \\
\cline{1-3} \cline{2-3}
 \multicolumn{1}{c|}{} & 6 & 17 \\
\cline{2-3}
\end{tabular}

\vskip0.5cm

\begin{tabular}{|c|cc|}
\hline
 $\frac{5}{6}$ & $\rightarrow$0 & $\rightarrow$1 \\
\hline \hline
 000 & 0 & 0 \\
\hline
 001 & 0 & 0 \\
 010 & 0 & 0 \\
 100 & 0 & 0 \\
\hline
 011 & 0 & 1 \\
 101 & 0 & 1 \\
 110 & 0 & 1 \\
\hline
 111 & 1 & 2 \\
\cline{1-3} \cline{2-3}
 \multicolumn{1}{c|}{} & 1 & 5 \\
\cline{2-3}
\end{tabular}
\begin{tabular}{|c|cc|}
\hline
 $\frac{6}{7}$ & $\rightarrow$0 & $\rightarrow$1 \\
\hline \hline
 000 & 0 & 0 \\
\hline
 001 & 0 & 0 \\
 010 & 0 & 0 \\
 100 & 0 & 0 \\
\hline
 011 & 0 & 1 \\
 101 & 0 & 1 \\
 110 & 0 & 1 \\
\hline
 111 & 1 & 3 \\
\cline{1-3} \cline{2-3}
 \multicolumn{1}{c|}{} & 1 & 6 \\
\cline{2-3}
\end{tabular}
\begin{tabular}{|c|cc|}
\hline
 $\frac{7}{8}$ & $\rightarrow$0 & $\rightarrow$1 \\
\hline \hline
 000 & 0 & 0 \\
\hline
 001 & 0 & 0 \\
 010 & 0 & 0 \\
 100 & 0 & 0 \\
\hline
 011 & 0 & 1 \\
 101 & 0 & 1 \\
 110 & 0 & 1 \\
\hline
 111 & 1 & 4 \\
\cline{1-3} \cline{2-3}
 \multicolumn{1}{c|}{} & 1 & 7 \\
\cline{2-3}
\end{tabular}
\begin{tabular}{|c|cc|}
\hline
 $1$ & $\rightarrow$0 & $\rightarrow$1 \\
\hline \hline
 000 & 0 & 0 \\
\hline
 001 & 0 & 0 \\
 010 & 0 & 0 \\
 100 & 0 & 0 \\
\hline
 011 & 0 & 0 \\
 101 & 0 & 0 \\
 110 & 0 & 0 \\
\hline
 111 & 0 & 1 \\
\cline{1-3} \cline{2-3}
 \multicolumn{1}{c|}{} & 0 & 1 \\
\cline{2-3}
\end{tabular}
\end{center}
\caption[Outcome statistics for the B-A-R model at various $m=3$
states.]{Outcome statistics for the B-A-R model at various $m=3$
states. The table shows the relative number of occurrence of each
outcome following every possible history bit-string for the states
characterized by $w_{\max}$ = $8/16$, $12/17$, $17/23$, $5/6$,
$6/7$, $7/8$, and $1$.  The parameter $\zeta$ labels groups of
histories as defined in Fig.~\ref{fig:hs_m3}.} \label{tab:tm_m3}
\end{table}

We have carried out detailed analysis of the outcomes following a
given history bit-string for $m=3$, and for each of the possible
states over the whole range of $L$, i.e., we obtain the chance of
getting 1 (or 0) for every 3-bit history by counting from the
outcome bit-string.  Table~\ref{tab:tm_m3} gives the {\em relative
numbers of occurrences} of each outcome for every history
bit-string.  For the state with $w_{\max} = 1/2 = 8/16$, for
example, the outcomes $0$ and $1$ occur with equal probability for
every history bit-string, as in the MG.  For the other states, the
results reveal several striking features. It turns out that
$w_{\max}$ is simply given by the {\em relative frequency} of an
outcome of $1$ in the outcome bit-string, which in turn is
governed by the resource level $L$. For example, a 0 to 1 ratio of
$5:12$ in the outcome bit-strings corresponds to the state with
$w_{\max}=12/17$. In Table~\ref{tab:tm_m3}, we have intentionally
grouped the history bit-strings into rows according to the label
$\zeta$ in Fig.~\ref{fig:hs_m3}.  We immediately notice that for
every possible state in the B-A-R model, the relative frequency of
each outcome is a property of the {\em group} of histories having
the same label $\zeta$ rather than the individual history
bit-string, i.e., all histories in a group have the same relative
fraction of a given outcome.  This observation is important in
understanding the dynamics in the history space for different
states in that it is no longer necessary to consider each of the
$2^{m}$ history bit-strings in the history space. Instead, it is
sufficient to consider the four groups of histories (for $m=3$) as
shown in Fig.~\ref{fig:hs_m3}(a).  Analysis of results for higher
values of $m$ show the same feature.

For the state characterized by $w_{\max}=1$, the outcome
bit-string is persistently $1$ and the path in the history space
is repeatedly 111$\rightarrow$1 (in the format of \emph{history}
$\rightarrow$ \emph{outcome}).  Therefore, the path simply
corresponds to an infinite number of loops around the history node
111.  Since the path is restricted to the $\zeta=0$ node, we will
also refer to this state as $\zeta_{\max}=0$ state. The system is
effectively frozen into one node in the history space. In this
case, there are effectively only \emph{two} kinds of strategies in
the whole strategy set, which differ by their predictions for the
particular history 111. The difference in predictions for the
other ($2^{m} -1$) history bit-strings become irrelevant.
Obviously, the ranking in the performance of the two effective
groups of strategies is such that the group of strategies that
suggest an action `1' for the history 111, outperforms the group
that suggests an action `0'. For a uniform initial distribution of
strategies, there are $N/2^{s}$ agents taking the action `0' and
$(1- 1/2^{s})N$ agents taking the action `1', since half of the
strategies predict 0 and half of them predict 1.  To sustain a
winning outcome of 1, the criterion is that the resource level $L$
should be higher than the number of agents taking the action `1'.
Therefore, we have for the state with $w_{\max} =1$ that
\begin{equation}
 \langle w \rangle =  1-\frac{1}{2^s} \label{eq:wmean_z0}
\end{equation}
and
\begin{equation}
 L_{\min} = \Big(1-\frac{1}{2^s}\Big)N. \label{eq:Lc_z0}
\end{equation}
These results are in agreement with numerical results. For $s=2$,
$\langle w \rangle=3/4$ for $L
> 3N/4$.  Note that Eqs.~\eqref{eq:wmean_z0} and
\eqref{eq:Lc_z0} are valid for {\em any} values of $m$.

Table~\ref{tab:tm_m3} shows that the states with $w_{\max} = 5/6$,
$6/7$, $7/8$ have very similar features in terms of the bit-string
statistics.  They differ only in the frequency of giving an
outcome of $1$ following the history of $111$.  Note that the
$\zeta = 2$ and $\zeta=3$ histories do not occur.  The results
imply that as the system evolves, the path in history space for
these states is restricted to the two groups of histories labeled
by $\zeta = 0$ and $\zeta=1$.  The statistics show that the
outcome bit-strings for the states with $w_{\max}=5/6$, $6/7$ and
$7/8$ exhibit only one $0$-bit in a period of $6$, $7$ and $8$
bits, respectively.  We refer to these states collectively as
$\zeta_{\max}=1$ states, since the portion of allowed history
space is bounded by the $\zeta=1$ histories. Graphically, the path
in history space consists of a few self-loops at the node 111,
i.e., from 111 to 111, then passing through the $\zeta=1$ group of
histories once and back to 111, as shown in
Fig.~\ref{fig:hs_m3}(b). The states with $w_{\max} = 1/2$,
$12/17$, $17/23$ involve the other groups of histories and exhibit
complicated looping among the histories. We refer to them
collectively as higher (i.e., $\zeta_{\max} > 1$) states.

\section{The $\zeta_{\max}=1$ states}
\label{sec:z1states}


The observed values of $w_{\max}$ for $\zeta_{\max} =1$ states can
also be derived by following the evolution of the strategy
performance ranking pattern as the game proceeds.  We summarize
the main ideas here.  Details can be found in Ref.\cite{my2004c}.
The major result is that for given $m$, the values of $w_{\max}$
in a $\zeta_{\max} =1$ state can only take on
\begin{equation}
 w_{\max} = \frac{m+n}{m+n+1}, \label{eq:wmax_z1}
\end{equation}
where $2 \leq n \leq m+1$.  Taking $m=3$ for example, we have $2
\leq n \leq 4$ and hence $n=2,3,4$. The values of $w_{\max}$ are
$5/6$, $6/7$, and $7/8$, exactly as observed in the numerical
simulations.

Eq.~(\ref{eq:wmax_z1}) is a result of the collective response of
the agents stemming from their strategy selection and
decision-making processes, which in turn are coupled to the
strategy ranking pattern.  To close the feedback mechanism, the
strategy ranking pattern must evolve in such a way so as to be
consistent with the path in the history space.  Recall that the
$\zeta_{\max}=1$ states are those in which the system only visits
the $\zeta = 0$ and $\zeta =1$ histories, as shown in
Fig.~\ref{fig:hs_m3}(b).  In this situation, only $(m+1)$ history
bits in a strategy are relevant, despite each strategy has $2^{m}$
entries, i.e., strategies that only differ in their predictions
for the histories which do {\em not} occur are now effectively
identical.  Therefore, many strategies have tied performance.

The key point is that a complete path of the $\zeta_{\max} =1$
states in history space corresponds to one in which there are $n$
turns around the $\zeta=0$ history, i.e., from $111 \rightarrow 1$
for $n$ turns, then breaks away to visit each of the $\zeta=1$
histories once and returns to the $\zeta=0$ history (see
Fig.~\ref{fig:hs_m3}(b)).  The value of $n$ that is consistent
with the condition $L < 3N/4$ turns out to be $2 \leq n \leq m+1$.
We have argued that for $L > 3N/4$, the system loops around the
$\zeta=0$ history indefinitely, since the number of agents
($3N/4$) persistently taking the option 1 is smaller than $L$, in
the large population limit.  For $L < 3N/4$, the situation is as
follows.  As the system starts to loop around the $\zeta=0$
history, the strategies start to split in performance, with the
group of strategies predicting $111 \rightarrow 1$ ($111
\rightarrow 0$) becomes increasingly better (worse).  There is an
overlap in performance between these two groups, as a result of
their predictions when the system passes through the $\zeta = 1$
histories.  As the number of loops at the $\zeta=0$ history
increases, the overlap in performance decreases and the number of
agents taking option 1 increases towards $3N/4$, as a result of
the strategy selection process.  The condition $L < 3N/4$, thus
imposes an upper bound on $n$.  When the number of turns at the
$\zeta =0$ history exceeds a certain value fixed by $L$, the
option 1 becomes the {\em losing} option and the system breaks
away to the $\zeta =1$ histories as the system makes transition
from the history 111 to the history 110.  The lower bound on $n$
is set by the restriction that the strategies in the group
predicting $111 \rightarrow 1$ must perform better on the average
than the group predicting $111 \rightarrow 0$, as $L > N/2$.  With
this understanding on the allowed paths in the history space, we
can evaluate the number of winners in each turn of the path and
obtain Eq.~(\ref{eq:wmax_z1}) for $w_{\max}$, with $n$ being the
number of loops that the system makes at the $\zeta=0$ history and
$2 \leq n \leq m+1$.

\section{B-A-R in Connected Populations}
\label{sec:BARCRG}

\begin{figure}[!htb]
\centerline{\epsfig{file=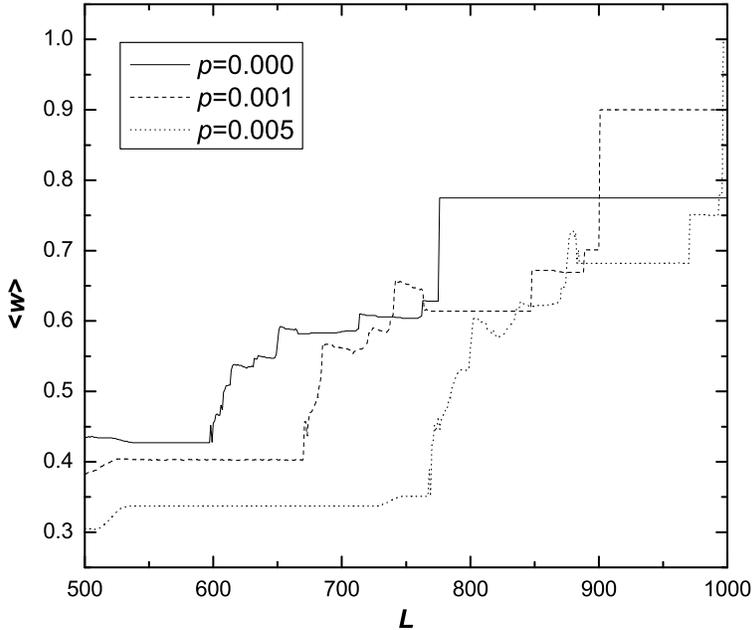,height=4in}}
\caption{The mean success rate $\langle w \rangle$ as a function
of resource level $L$ for $N=1001$, $s=2$, and $m=1$.  Different
lines denote different values of $p$ in a classical random graph
(CRG).  Data for different $L$ are obtained from the same
connected population and distribution of strategies among the
agents.} \label{fig:wmCRG_m3s2}
\end{figure}

\begin{figure}[!htb]
\centerline{\epsfig{file=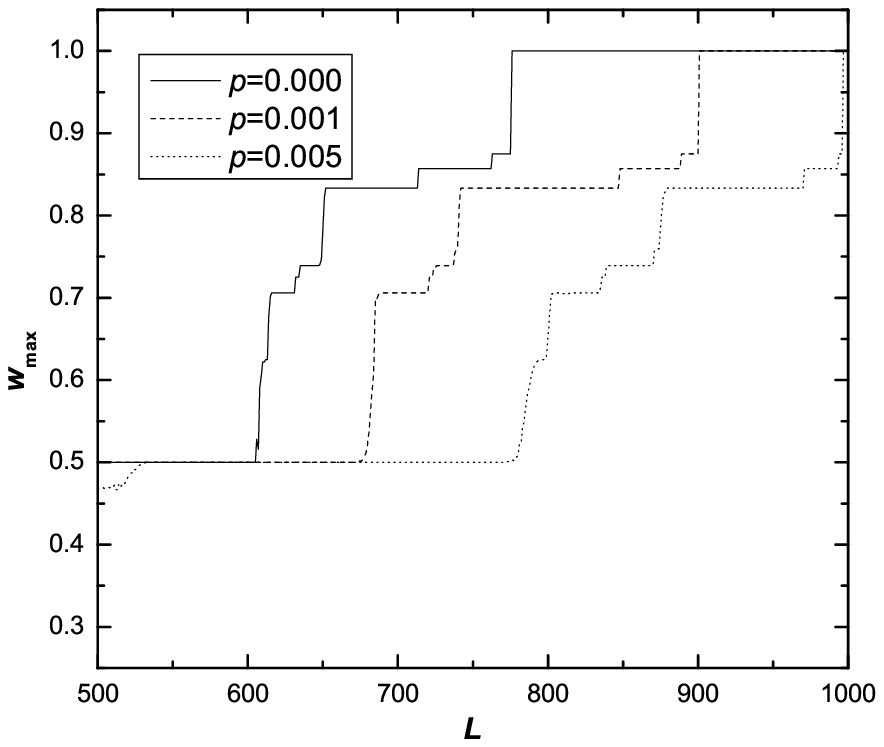,height=4in}}
\caption{The highest success win rate $w_{max}$ among the agents
as a function of resource level $L$ for $N=1001$, $s=2$, and
$m=1$. Different lines denote different values of $p$ in a
classical random graph (CRG).  Data are obtained from the same
configurations as in Figure~\ref{fig:wmCRG_m3s2}.}
\label{fig:wxCRG_m3s2}
\end{figure}

We have also studied the B-A-R model in a connected population.
The agents are assumed to be connected in the form of a classical
random graph (CRG), i.e., random network.  In a CRG, each node
(representing an agent here) has a probability $p$ of connecting
to another node.  The agents interact in the following way. When
an agent is linked to other agents, he checks if the strategies of
his connected neighbours perform better than his own strategies.
If so, he will choose the one with the highest VP among his
neigbhours and then use it for decision in that turn. If not, he
will use his own best strategy.  We have carried out detailed
numerical calculations of $\langle w \rangle$ and $w_{max}$.
Figures~\ref{fig:wmCRG_m3s2} and \ref{fig:wxCRG_m3s2} show how
$\langle w \rangle$ and $w_{max}$ depend on $L$ for various
connecting probability $p$, with the $p=0$ results corresponding
to those in a non-networked population.

We observe that the features of plateaux-and-jumps persist in
networked populations.  There are differences in the details.
While $w_{\max}$ takes on the same set of values, the threshold in
the resource level ($L_{\min}$) to sustain a particular level of
success rate is found to be higher in a networked population than
in a non-networked one, as shown in Fig.~\ref{fig:wmCRG_m3s2} and
Fig.~\ref{fig:wxCRG_m3s2}.  For a given resource level, allowing
agents to share information on strategy performance may actually
worsen the global performance of the system.  This is
particularly clear if we inspect the results of $\langle w
\rangle$ in the range of $0.5 < L/N < 0.75$ in
Fig.~\ref{fig:wmCRG_m3s2}.  The addition of a small number of
links will lower $\langle w \rangle$.  This behaviour is
consistent with the crowd-anticrowd theory
\cite{Johnson2003b,crowd1,crowd2} in that the links enhance the
formation of crowds.  The situation can be quite different in a
high resource limit $L/N \approx 1$, where a small number of
links may actually be beneficial. The reason is that for high
resource level, some resource is left unused as $N/4$ agents will
not have access to a strategy that predicts the winning option.
Using the links, some of these agents switch from losers to
winners and thus enhance the overall performance of the population
\cite{Gourley2004a}.

\section{Summary}
\label{sec:Discussion}

We have studied numerically and analytically the effects of a
varying resource level $L$ on the success rate of the agents in a
competing population within the B-A-R model.  We found that the
system passes through different states, characterized either by
the mean success rate $\langle w \rangle$ or by the highest
success rate in the population $w_{\max}$, as $L$ decreases from
the high resource level limit. Transitions between these states
occur at specific values of the resource level.  We found that
different states correspond to different paths covering a subspace
within the whole history space.  Just below the high resource
level is a range of $L$ that gives $m$ states corresponding to the
fractions $w_{\max} = (m+n)/(m+n+1)$, with $2 \leq n \leq m+1$.
This result is in excellent agreement with that obtained by
numerical simulations. The paths of these states in the history
space are restricted to those $m$-bit histories with at most
one-bit of 0 and with $n$ loops around the 111\ldots history. This
result is derived by considering the coupling of the restricted
history subspace that the system visits, the strategy performance
ranking pattern, and the strategy selection process.  While our
analysis can also be applied to the $\zeta_{\max}>1$ states, the
dynamics and the results are too complicated to be included here.

Our analysis also serves to illustrate the sensitivity within
multi-agent models of competing populations, to tunable
parameters.  By tuning an external parameter, which we take as the
resource level in the present work, the system is driven through
different paths in the history space which can be regarded as a
`phase space' of the system. The feedback mechanism, which is
built-in through the decision making process and the evaluation of
the performance of the strategies, makes the system highly
sensitive to the resource level in terms of which states the
system decides to settle in or around. These features are quite
generally found in a wide range of complex systems.  The ideas in
the analysis carried out in the present work, while specific to
the B-A-R model used, should also be applicable to other models of
complex systems.

While we have focused on the B-A-R model in our discussions, many
of the underlying ideas are general to a wider class of complex
systems.  For example, one may regard the resource level $L$ as a
handle in controlling a driving force in the system.  With
$L=N/2$, i.e., in the MG, and the random initial distribution of
strategies and random initial history, the system is allowed to
diffuse from an initial node in the history space to visit all the
possible histories.  A deviation of the resource level from $N/2$
acts like a driving force in the history space.  Thus, there is
always a competition between diffusive and driven behaviour,
resulting in the non-trivial behaviour in the B-A-R model and its
variations.  For this reason, the present B-A-R system provides a
fascinating laboratory for studying correlated, non-Markovian
diffusion on a non-trivial network (i.e., history space).

\acknowledgments{Work at CUHK was supported in part by a grant from the
Research Grants Council of the Hong Kong SAR Government.  Sonic H. Y. Chan
acknowledges the support from CUHK for attending WEHIA 2005.}

\end{document}